# The Relativistic Foundations of Synchrotron Radiation

**G. Margaritondo[a]* and J. Rafelski[b]**

[a]Ecole Polytechnique Fédérale de Lausanne (EPFL), Switzerland,

[b]Department of Physics, The University of Arizona, Tucson, AZ, USA

*Correspondence e-mail: giorgio.margaritondo@epfl.ch

Special Relativity (SR) determines the properties of synchrotron radiation, but the corresponding mechanisms are frequently misunderstood. Time dilation is often invoked among the causes, whereas its role would violate the principles of SR. We show that the correct explanation of the synchrotron radiation properties is provided by a combination of the Doppler shift, not dependent on time dilation effects, contrary to a common belief, and of the Lorentz transformation into the particle reference frame of the electromagnetic field of the emission-inducing device, also with no contribution from time dilation. We close by reminding the reader that much if not all of our argument has been available since the inception of SR, a research discipline of its own standing.

Synchrotron sources are one of the main practical applications of Einstein's SR (Einstein, 1905-1). Paradoxically, however, the relativistic effects that determine the detected radiation features are typically presented in a misleading way.

Consider the $2\gamma^2$-term ubiquitously found in the spectral equations of synchrotron radiation (Margaritondo, 2002). For example, the first-harmonic emitted wavelength of an undulator is approximately proportional to $L/(2\gamma^2)$, $L$ being the undulator period, and the photon energy bandwidth of a bending magnet is proportional to $\approx 2\gamma^2 B$, with $B$ = magnetic field strength.

The $\approx 2\gamma^2$-term actually originates from the product of two factors, $\approx 2\gamma$ and $\gamma$, related to two different relativistic phenomena. Let us consider what they are.

The $\approx 2\gamma$-factor is due to the longitudinal Doppler shift occurring when the emission of the moving electrons is observed in the laboratory frame R. Unfortunately, the Doppler shift equations are frequently derived using time dilation, which is fundamentally incorrect (Rafelski, 2017). The $\gamma$-factor is also often attributed to time dilation, likewise a conceptual mistake.

Consider first the Doppler shift, using the geometry that is relevant to synchrotron radiation: the source (the electron) has a longitudinal component of the velocity directly towards the observer, with magnitude $u \approx c$. Before treating the effects of this movement, we must accurately define the "electron" reference frame R'. This is *not* the frame moving with the electron (which would not be inertial since the electron is accelerated in the transverse direction). Rather, it is the frame moving with a constant velocity that is equal to the instantaneous longitudinal velocity component of the electron. In this frame, the electron has zero longitudinal speed and acceleration, but it can be transversely accelerated and emit radiation.

To explain the relativistic Doppler shift for electromagnetic waves in vacuum, a common but conceptually incorrect (Rafelski, 2017) approach - whose historical origins are discussed below - adds time dilation to the non-relativistic Doppler effect, e.g., for sound. The starting equation is, therefore:

$$v = \frac{v'}{\left(1 - \frac{u}{c}\right)},$$

where $v$ and $v'$ are the frequencies in R and R', and $c$ is the speed of the wave. Then, the equation is modified for time dilation. The logic is that the time intervals measured by clocks are increased by the factor $\gamma = (1 - u^2/c^2)^{-1/2}$ when going from R' to R, so the frequencies should be reciprocally increased by the same factor from R to R', and:

$$v = \frac{v'}{\left(1 - \frac{u}{c}\right)} \sqrt{1 - \frac{u^2}{c^2}} = \sqrt{\frac{1 + \frac{u}{c}}{1 - \frac{u}{c}}} \, v',$$

which in the limit $u \approx c$ becomes:

$$v = \frac{1 + \frac{u}{c}}{\sqrt{1 - \frac{u^2}{c^2}}} \approx 2\gamma v'.$$

Paradoxically, these equations are correct (for a special line of sight) but the logic above is not, as explained in (Rafelski, 2017, p. 172): "*To understand the absurdity of claims that the SR-Doppler [Special Relativity-Doppler] effect is created at the source due to time dilation, the reader should consider not two but three different observers, e.g. two different travelers and the laboratory source. There are three different relative velocities. There are three different SR-Doppler shifts observed that must be relative and reciprocal. The only way that this can be true is that the Doppler shift is created by the method of observation and not by state of motion of the source*".

Regrettably, the incorrect time dilation explanation is what students frequently learn, notably from Wikipedia (https://en.wikipedia.org/wiki/Relativistic_Doppler_effect),

which states "*The relativistic Doppler effect is different from the non-relativistic Doppler effect as the equations include the time dilation effect of special relativity*". Actually, the origin of the misconception is much older: it can be traced to the SR text (Resnick 1968) and subsequently percolated into popular teaching books like the "Halliday-Resnick-Krane" physics manuals (Halliday et al. 2002).

Resnick presented the Doppler shift and the line-of-sight aberration following von Laue's classic relativity text in German (von Laue 1960), which included the following comment: "*Die Wurzel* $\sqrt{1-\beta^2}$ *in 14.8 welche den quadratischen Doppler-Effekt bedingt, stammt, wie man an Hand der Rechnung Zurüverfolgt, aus dem Nenner* $\sqrt{1-\beta^2}$ *der Transformation (4.7). Da wir aus demselben Nenner in § 5a auf die Zeitdilatation schlossen, finden wir auch hier den Zusammenhang zwischen dem quadratischen Effekt und der Verlangsamung des Uhrenganges…*" A correct paraphrased translation is "*...both the Doppler γ-factor and time dilation γ-factor originate in the γ-factor of the Lorentz transformation*". Instead, Resnick's version was: "*It is instructive to note that the transverse Doppler effect has a simple time-dilation interpretation. ..... . The transverse Doppler effect is another physical example confirming the relativistic time dilation.*" The use of the words "interpretation" and "confirming" is the conceptual mistake that influenced the subsequent teaching of the Doppler effect – and synchrotron radiation – over one-half century.

In essence, the above derivation of the Doppler shift and its time dilation argument live in a hypothetical (and wrong) world filled with **material (or absolute) aether**. Consider the general Doppler effect for sound: the frequency of emitted waves depends on the relative velocity of the source with respect to air; there is a second shift if the observer also moves with respect to air. The time-dilation argument for the Doppler shift of light uses the same logic: the emitted frequency is modified by the relative motion of the source with respect to the carrier of emitted radiation, the material aether. This thinking violates of course the principle of relativity by *de facto* assuming that the modification is with respect to an absolute aether -- whereas Einstein (Einstein, 1905-1) saved Galileo's principle of relativity by recognizing that it is the observer who creates in the measurement process the Doppler shift for light.

There is, in fact, no need to incorrectly invoke time dilation when deriving the Doppler shift, as demonstrated in (Rafelski, 2017). Consider for simplicity a plane and linearly polarized wave traveling in the longitudinal direction z, whose (transverse) magnetic field magnitude can be written in the R'-frame as:

$$B' = B_o' \sin\left[2\pi\left(\frac{z'}{\lambda'} - v't'\right)\right],$$

where $\lambda'$ is the wavelength and $v'$ is the frequency. The phase factor of this wave must be Lorentz-invariant, otherwise one could use phase-related effects to detect the inertial motion of R' with respect to R (or vice-versa), violating the principles of SR (Einstein, 1905-1). Therefore:

$$\frac{z'}{\lambda'} - v't' = \frac{z}{\lambda} - vt.$$

Using the wave relations $\lambda' = c/v'$ and $\lambda = c/v$ as well as the Lorentz transformations for the longitudinal position and for the time, this equation gives:

$$\left(\frac{z'}{c} - t'\right)v' = \left(\frac{z}{c} - t\right)v;$$

$$\left[\frac{\gamma(z-ut)}{c} - \gamma\left(t - \frac{uz}{c^2}\right)\right]v' = \left(\frac{z}{c} - t\right)v;$$

$$\gamma\left(1 + \frac{u}{c}\right)\left(\frac{z}{c} - t\right)v' = \left(\frac{z}{c} - t\right)v,$$

and, therefore:

$$v = \frac{\left(1 + \frac{u}{c}\right)}{\sqrt{1 - \frac{u^2}{c^2}}} v' = \sqrt{\frac{1 + \frac{u}{c}}{1 - \frac{u}{c}}} v',$$

the same relativistic Doppler equation as above, but derived here without invoking time dilation.

An alternate way to achieve the same objective is to Lorentz-transform the energy carried by hypothetical "energy particles" associated with the wave (Einstein's "photon hypothesis" (Einstein, 1905-2)). The procedure, reported in (Margaritondo, 1995), shows that:

$$P = \sqrt{\frac{1+\frac{u}{c}}{1-\frac{u}{c}}} P',$$

where $P$ and $P'$ are the particle energies in the two frames. Thus, assuming that frequencies and particle energies are linearly related, $P = h\nu$ and and $P' = h\nu'$, one obtains the Doppler shift equation, again without using time dilation. The bonus of this approach is to shed light on an often-neglected link between relativity and quantum electromagnetism.

Before concluding the discussion of the Doppler effect, one should note that the above discussion is only valid if one considers the transverse motion of the electron as negligible with respect to the relativistic longitudinal motion. Strictly speaking, this is not correct and one should use instead the general theory of the Doppler effect valid for all directions of the source motion. This does not change the fact that invoking time dilation is incorrect. But the transverse motion does have important effects. For example, by increasing its amplitude one decreases the Doppler shift in the longitudinal direction -- an effect that is widely exploited [2] to control the undulator spectral emission by changing its magnetic field strength.

We shall now discuss the second relativistic effect in synchrotron radiation, which causes the $\gamma$ term in the $\approx 2\gamma^2$ factor. Here too, one should avoid the erroneous use of time dilation since, quoting again (Rafelski, 2017, p. 172): "*Light travels in space in a way that must be completely independent of the state of motion of the source; there cannot be an effect of motion of the source that can be ascribed to properties of* *emitted light*".

To understand this point, note that in all synchrotron radiation sources the emission is triggered by a magnetic device such as a bending dipole, an undulator or a wiggler (Margaritondo, 2002) -- which produces the transverse acceleration of the electron. Consider for example the (transverse) magnetic field magnitude of an undulator, assuming a simple oscillating form in the R-frame:

$$B_U = B_{Uo} \sin\left[2\pi\left(\frac{z}{L}\right)\right],$$

where *L* is again the undulator period. After the Lorentz transformation to the "electron" R'-frame, this field magnitude becomes:

$$B_U' = \gamma B_{Uo} \sin\left[2\pi\gamma\left(\frac{z'+ut'}{L}\right)\right].$$

Furthermore, the transformation adds to the magnetic field a transverse electric field, perpendicular to it and with magnitude:

$$E_U' = \gamma\left(\frac{u}{c^2}\right) B_U'.$$

For the relativistic limit *u* ≈ *c*, these last two equations become, approximately:

$$B_U' = \gamma B_{Uo} \sin\left[2\pi\gamma\left(\frac{z'+ct'}{L}\right)\right];$$

$$E_U' = \gamma\left(\frac{1}{c}\right) B_U'.$$

These can be recognized (Margaritondo, 2002) as the equations of an electromagnetic wave moving in the negative longitudinal direction towards the electron – with wavelength *L*/γ and frequency γ*c*/*L*. The electric field of this "wave" acts on the electron whereas the magnetic field does not, since in R' the longitudinal velocity is zero and there is no Lorentz force.

The interaction with the electron causes the elastic back scattering of the "wave" in the R'-frame: this is the mechanism that generates synchrotron radiation (Margaritondo, 2002). The wavelength and the frequency of the back scattered wave in R' are the same as those of the original wave, $L/\gamma$ and $\gamma c/L$. Therefore, the emitted frequency includes the factor $\gamma$ that contributes to the $\approx 2\gamma^2$ term, which is thus explained by the Lorentz field transformation without involving time dilation.

Such a result could be re-interpreted by noting that the $L/\gamma$ value of the wavelength corresponds to the Lorentz-contracted undulator period. This point of view is conceptually correct: the field transformation that defines the emission-triggering object in R' (the undulator "wave" that is back scattered) does include the longitudinal contraction.

One could likewise be tempted to re-interpret the factor $\gamma$ in the emitted frequency in R' as a reverse effect of the time dilation affecting the clocks in R' when observed in the laboratory frame. However, this interpretation falls in the same kind of conceptual mistake discussed above (Rafelski, 2017).

This is a subtle but very important point: an observer in the frame R' would not know the state of relative motion of the laboratory observer that will eventually detect the radiation – who *may not even exist*. Therefore, one cannot expect this relative state of motion to influence the emitted waves in R'. In a synchrotron radiation source, what matters instead is the state of motion of the emission-triggering magnetic device with respect to the electron, which only coincidentally is the same as that of the potential laboratory observer of the radiation. The relative motion of the magnetic device (and not that of the observer) with respect to the electron, treated with the Lorentz transformation of the magnetic field, defines what *is* the electromagnetic system in R' that interacts with the electron to produce synchrotron radiation. This has nothing to do with the relative state of motion of the possible future laboratory observer, and in particular with the time dilation between the electron frame and this observer's frame.

To further clarify this point, we should note that having the same reference frame for the undulator and for the laboratory observer is only a practical matter. But there is no fundamental reason for not imagining the undulator and the laboratory observer moving in different ways with respect to the electron. In that case, the $2\gamma^2$ term would be replaced by $2\gamma_{eO}\gamma_{eB}$, where the two factors $\gamma_{eO}$ and $\gamma_{eB}$ correspond to the two distinct relative motions, electron-observer and electron-undulator.

The $\gamma_{eB}$ term enhances the strength of the transverse electromagnetic field in the R' ("electron") frame and changes its periodicity. In other words, it determines what is in the same frame of the "apparatus" causing the electron to emit radiation. The $\gamma_{eO}$ term is instead the relativistic gamma factor of an observer measuring the emitted radiation that already left the apparatus and is on its way. This line of reasoning clarifies why the state of motion of the final observer with respect to the source cannot influence the emission properties in R'.

Similar conclusions are valid for other types of synchrotron sources. For bending magnets, for example, the Lorentz transformation of the field gives in the R'-frame a transverse electric field with a factor of $\gamma$ -- which then is found in the emitted wavelength and frequency. As for the case of undulators, the Lorentz-transformed field fully defines what really *is*, in R', the emission-inducing electromagnetic device.

In conclusion, the $\approx 2\gamma^2$ term in the spectral equations of synchrotron radiation is the result of two different relativistic effects: the longitudinal Doppler effect and the Lorentz transformation of the magnetic field of the device that induces the emission. Neither one of them should be attributed, even in part, to the time dilation affecting the clocks in the "electron" frame when seen in the observer (laboratory) frame – as is frequently and unfortunately done.

Note that nothing we described above is new. Our ideas were already present in Einstein's work of 1905 (Einstein, 1905-1). So, closing this discussion we ask the question: how could it be that after more than one century we are correcting some of the common views about synchrotron radiation phenomena? We discussed already how the wrong understanding of relativity phenomena started to infiltrate textbooks about 50 years ago. One should not neglect, however, another serious problem: SR is often not taught as a separate discipline, and "General Relativity" tends to be seen as its teaching home. This feeds conceptual misunderstandings about SR, as its effects are often confused with modifications of space-time. The ensuing misconceptions specifically affect synchrotron radiation phenomena, but are not limited to them: the Lorentz contraction of material bodies is another example.

The clarification of the origin of the synchrotron radiation factor $2\gamma^2$ described in this article hopefully assures that modeling and technological improvements are in the future well aligned with foundational principles of SR. Our discussion of the origins of misleading views about the principles of relativity should help the readers – in particular teachers and students - to critically reject texts and web resources that propagate conceptual misunderstandings about the Doppler effect and about synchrotron radiation in general.

**Acknowledgments**

This work was supported by the Ecole Polytechnique Fédérale de Lausanne (EPFL) and by its Center for Biomedical Imaging (CIBM).

**References**

Einstein, A. (1905-1). *Ann. Phys.* **17**, 891-921.

Einstein, A. (1905-2). *Ann. Phys.* **17**, 132-148.

Margaritondo, G. (2002). *Elements of Synchrotron Light for Biology, Chemistry, and Medical Research*, New York: Oxford.

Margaritondo, G. (1995). Eur. J. Phys. **16**, 169-171.

Rafelski, J. (2017) *Relativity Matters: From Einstein's EMC2 to Laser Particle Acceleration and Quark-Gluon Plasma*, Berlin: Springer.

Resnick, R. (1968). *Introduction to Special Relativity,* New York: J. Wiley, ppp. 90-91.

von Laue, M. (1960) *Die Relativitätstheorie,* Braunschweig: Vieweg; 7. durchgesehene Auflage pp. 101-2.

Halliday, D., Resnick, R., Krane, K. S. (2002) *Physics*, New York: J. Wiley see in this 4[th] edition Volume 2 extended pp. 897-8.